\documentclass[english]{article}
\usepackage[T1]{fontenc}
\usepackage[latin9]{inputenc}
\usepackage{color}
\usepackage{amsmath}
\usepackage{amssymb}
\usepackage{esint}

\makeatletter
\newcommand{\lyxaddress}[1]{
\par {\raggedright #1
\vspace{1.4em}
\noindent\par}
}

\makeatother

\usepackage{babel}
\begin{document}

\title{\textcolor{blue}{\huge Newtonian Semiclassical Gravity in the Ghirardi-Rimini-Weber
Theory with Matter Density Ontology}}

\author{Maaneli Derakhshani%
\thanks{Emails: maanelid@yahoo.com and maaneli.derakhshani@unl.edu.%
}%
\thanks{Tel.: +1 917 257 0311.%
}}

\maketitle

\lyxaddress{\begin{center}
\textit{\large Department of Physics \& Astronomy, Clemson University,
Clemson, SC 29634, USA}%
\footnote{Present address: \textit{Department of Physics \& Astronomy, University
of Nebraska-Lincoln, Theodore P. Jorgensen Hall, Lincoln, NE 68588,
USA. }%
}
\par\end{center}}
\begin{abstract}
We propose a Newtonian semiclassical gravity theory based on the GRW
collapse theory with matter density ontology (GRWm), which we term
GRWmN. The theory is proposed because, as we show, the standard Newtonian
semiclassical gravity theory based on the Schroedinger-Newton equations
does not have a consistent Born rule probability interpretation for
gravitationally self-interacting particles and implies gravitational
cat states for macroscopic mass superpositions. By contrast, we show
that GRWmN has a consistent statistical description of gravitationally
self-interacting particles and adequately suppresses the cat states
for macroscopic superpositions. Two possible routes to experimentally
testing GRWmN are also considered. We conclude with a discussion of
possible variants of GRWmN, what a general relativistic extension
would involve, and various objections that might be raised against
semiclassical gravity theories like GRWmN.
\end{abstract}

\lyxaddress{\begin{center}
\par\end{center}}

\section{Introduction}

The problem of how to consistently couple a classical gravitational
field to quantized matter was first addressed by Moeller \cite{Moeller1962}
and Rosenfeld \cite{Rosenfeld1963} in the 1960's, who proposed the
semiclassical Einstein equation (also called the {}``Moeller-Rosenfeld''
equation)

\begin{equation}
G_{nm}=\frac{8\pi G}{c^{4}}<\psi|\hat{T}_{nm}|\psi>,
\end{equation}
where $<\psi|\hat{T}_{nm}|\psi>$ is interpreted as the (either second
quantized or first quantized) quantum expectation value of the stress-energy-momentum
tensor operator ($\hat{T}_{nm}[\hat{\phi},g]$ in the second quantized
case and $\hat{T}_{nm}(\hat{x},g)$ in the first quantized case).
Motivated from standard quantum mechanics, this is argued \cite{Carlip2008,Kiefer2012}
to be the only consistent way of incorporating a quantum description
into the right hand side of (1) while keeping the left hand side a
classical field%
\footnote{The other possible approaches on might try are 1) to equate the left
hand side with the stress-energy tensor operator, and 2) to make an
eigenvalue equation out of the Einstein equation. The first possibility
doesn't make physical or mathematical sense because one equates a
c-number with an operator, while the second possibility fails because
the components of the stress-energy tensor don't commute and cannot
be simultaneously diagonalized. %
}. Equation (1) can also be formally derived from the semiclassical
approximation of the Wheeler-deWitt equation in canonical quantum
gravity, for N quantum matter fields interacting with the quantized
gravitational field as N $\rightarrow\infty$ \cite{Kiefer2012,B.L.Hu2008};
it can also be motivated by various (non-string theoretic) approaches
to emergent gravity \cite{Volovik2013,S.Finazzi2012,Hu2009,C.Barcelo2001}. 

It is noteworthy that (1) is the basis for several major results in
theoretical astrophysics in the 70's and 80's - Hawking radiation
from black holes, the cosmological perturbations generated in cosmic
inflation, particle pair production on expanding spacetimes, the creation
of naked black hole singularities, traversable wormhole solutions,
and warp drive spacetimes, to name a few \cite{Ford2005,N.D.Birrell1982}.
In the 90's and 2000's, a semiclassical gravity theory known as {}``stochastic
gravity'' \cite{B.L.Hu2008} was proposed as a theoretical bridge
between (1) and the as-yet-unknown theory of quantum gravity, and
has led to new theoretical predictions for the aforementioned astrophysical
phenomena. And most recently, a variant of the semiclassical gravity
theory based on (1), known as {}``gravitational-aether'' theory,
has been proposed as an observationally testable solution to the old
cosmological constant problem \cite{S.Aslanbeigi2011} and as a novel
solution to the endpoint of gravitational collapse in black holes
\cite{M.Saravani2012}. In addition to these general relativistic
results, a nonrelativistic approximation of (1) known as the Schroedinger-Newton
(SN) equations has been used by several researchers \cite{Carlip2008,Salzman2005,Penrose1996,Penrose1998,Diosi1984,D.Giulini2013,Meter2011,H.Yang2013,DomenicoGiulini2011,Diosi2012A,Diosi2012B,Adler2007}
to predict nonlinear semiclassical gravitational effects that could
be observable in future experiments with macro-molecular interferometry
\cite{Carlip2008,Salzman2005,D.Giulini2013,Meter2011,DomenicoGiulini2011,Adler2007,K.Hornberger2012,P.Asenbaum(2013)}
and other macroscopic/mesoscopic quantum systems \cite{H.Yang2013,Diosi2012B,Adler2007,W.Marshall2003}.
Thus, research on semiclassical gravity theories based on (1) is a
highly active area, with potentially significant implications for
theoretical astrophysics, cosmology, and quantum gravity phenomenology.

However, in spite of all these theoretical motivations and applications,
the formulation of semiclassical gravity based on (1) has serious
difficulties in its foundation, among them being the lack of a consistent
Born rule probability interpretation for gravitationally self-interacting
nonrelativistic particles \cite{Adler2007}, and the well-known gravitational
`cat-state' solutions \cite{Kiefer2012,Ford2005,Penrose1996,Diosi1984}
which are known to be experimentally ruled-out \cite{D.N.Page1981}.
Both problems have been previously discussed in the context of the
SN equations \cite{Kiefer2012,Ford2005,Penrose1996,Diosi1984,Adler2007},
and the primary aim of this paper is to propose a modification of
the SN equations in which the collapse dynamics of Ghirardi-Rimini-Weber
(GRW) \cite{G.C.Ghirardi1986,Allori2012} is used to give a consistent
statistical description of gravitationally self-interacting particles,
and to adequately suppress the cat state solutions for macroscopic
superpositions. 

We begin our paper by reviewing the arguments showing inconsistency
with the Born rule interpretation, and suggest what the correct physical
interpretation of the SN equations should be. We then review the existence
of macroscopic gravitational cat-state solutions by way of a simple
example and show that such solutions, coupled with the correct physical
interpretation of the SN equations, imply that the SN equations do
not correctly describe the semiclassical gravitational field for macroscopic
matter distributions. Next we develop the GRW collapse modification
of the SN equations, showing how it leads to a consistent statistical
description of gravitationally self-interacting particles and adequately
suppresses cat states for macroscopic superpositions. We then discuss
experimental possibilities for testing our GRW modification of the
SN equations, by re-examining two physical scenarios within which
the SN equations have been recently suggested to make new predictions
within experimentally testable reach. We conclude the paper with a
discussion of possible variants of our proposed GRW modification,
a discussion of what would be involved in a general relativistic extension,
and a discussion of possible objections that might be raised against
semiclassical gravity theories more generally.

\section{Inconsistency with the Born rule interpretation}

To illustrate the inconsistency with the Born rule interpretation
for gravitationally self-interacting particles, we first take the
Newtonian limit of (1), making the simplifying assumption that $|\psi>$
is a first-quantized wavefunction and $\hat{T}_{nm}=\hat{T}_{nm}(\hat{x},g)$
so that (1) specifically corresponds to the Einstein-Klein-Gordon
system \cite{D.Giulini2012,F.SiddharthaGuzman2003}. With the approximations
$g_{nm}=\eta_{nm}+h_{nm}$, $|T^{ij}|/T^{00}<<1$, and $v<<c,$ it
can readily be shown \cite{D.Giulini2012,F.SiddharthaGuzman2003}
that (1) reduces to the semiclassical Newton-Poisson equation,

\begin{equation}
\nabla^{2}V(x,t)=4\pi Gm|\psi(x,t)|^{2},
\end{equation}
with solution 
\begin{equation}
V(x,t)=-G\int\frac{m|\psi(x',t)|^{2}}{|x-x'|}d^{3}x',
\end{equation}
and $\psi$ satisfying the nonlinear integro-differential Schroedinger
equation,

\begin{equation}
i\hbar\partial_{t}\psi(x,t)=-\frac{\hbar^{2}}{2m}\nabla^{2}\psi(x,t)+mV(x,t)\psi(x,t).
\end{equation}
The N-body generalizations%
\footnote{It should be noted that our assumption that the SN equations have
a valid N-body generalization presumes a particular approach to (1);
namely, the approach of Moeller and Rosenfeld, Kibble and Randjbar-Daemi
\cite{T.W.Kibble1980}, and Boughn \cite{Boughn2009}, who consider
(1) as a fundamental description of gravity for a single particle
or field as well as for many particles or fields. By contrast, in
canonical quantum gravity, (1) is valid only as a single-body mean-field
equation, when one has N matter fields or particles interacting with
the quantized gravitational field as N $\rightarrow\infty$ \cite{Kiefer2012,B.L.Hu2008}.
Nevertheless, this difference will not affect the validity of our
subsequent analyses, which will apply to both the single-body and
N-body versions of the SN equations. %
} (ignoring the interaction potential term for simplicity) are as follows:

\begin{equation}
\nabla^{2}V(x,t)=4\pi G\int dx'_{1}...dx'_{N}|\psi(x_{1}'...x_{N}',t)|^{2}\underset{i=1}{\overset{N}{\sum}}m_{i}\delta^{3}(x-x'_{i}),
\end{equation}
and

\begin{equation}
i\hbar\partial_{t}\psi(x_{1}...x_{N},t)=-\underset{i=1}{\overset{N}{\sum}}\frac{\hbar^{2}}{2m_{i}}\nabla_{i}^{2}\psi(x_{1}...x_{N},t)+\underset{i=1}{\overset{N}{\sum}}m_{i}V(x_{i},t)\psi(x_{1}...x_{N},t),
\end{equation}
with the solution to (4) given by

\begin{equation}
V(x_{i},t)=-G\sum_{j=1}^{N}\int\frac{m_{j}|\psi(x'_{1}...x'_{N},t)|^{2}}{|x_{i}-x_{j}'|}dx'_{1}...dx'_{N}.
\end{equation}

The coupled equations defined by (2)-(4) or (5)-(6) are the single
and many particle SN equations respectively \cite{Kiefer2012,Diosi1984}.
Following Adler \cite{Adler2007}, we now make the independent particle
approximation of (5)-(7) by introducing the ansatz,

\begin{equation}
\psi(x_{1}...x_{N},t)=\prod_{r=1}^{N}\psi_{r}(x_{r},t),
\end{equation}
with each single particle $\psi_{r}$ normalized through

\begin{equation}
\int d^{3}x_{r}|\psi_{r}(x_{r},t)|^{2}=1.
\end{equation}
Using (8), Adler shows in a straightforward derivation that (6)-(7)
take the form

\begin{equation}
i\hbar\partial_{t}\psi_{s}(x_{s},t)=-\frac{\hbar^{2}}{2m_{s}}\nabla_{s}^{2}\psi_{s}(x_{s},t)+U(x_{s},t)\psi_{s}(x_{s},t),
\end{equation}
where 

\begin{equation}
U(x_{s},t)=-G\sum_{u=1}^{N}\int d^{3}x'_{u}\frac{m_{s}m_{u}|\psi_{u}(x'_{u},t)|^{2}}{|x_{s}-x'_{u}|}.
\end{equation}
Note that although (10)-(11) has the same structure as the time-dependent
single particle Schroedinger equation that one gets by treating the
Newtonian inter-particle potential in the Hartree approximation, it
differs in that, for $u=s$, it includes a self-interaction term of
the form

\begin{equation}
-G\int d^{3}x'_{s}\frac{m_{s}^{2}|\psi_{s}(x'_{s},t)|^{2}}{|x_{s}-x'_{s}|}.
\end{equation}
To see that such a term does not have a consistent Born rule interpretation,
we should first consider a term with $u\neq s$, 

\begin{equation}
-G\int d^{3}x'_{u}\frac{m_{u}m_{s}|\psi_{u}(x'_{u},t)|^{2}}{|x_{s}-x'_{u}|}.
\end{equation}
According to the Born rule interpretation, this would say that the
gravitational potential felt by particle $s$ at coordinate $x_{s}$,
as a result of the presence of particle $u$ at $x'_{u}$, is the
Newtonian potential $-Gm_{u}m_{s}/|x_{s}-x'_{u}|$ weighted by the
probability $|\psi_{u}(x'_{u},t)|^{2}$ of finding the particle $u$
at $x'_{u}$. However, as Adler notes, this interpretation does not
apply to the case of $u=s$, since when particle $s$ is at $x_{s}$,
the probability of simultaneously finding it at $x'_{s}$ is clearly
zero. He also shows this in terms of projection operators - for the
case $u\neq s$, we have that $P_{u}(x'_{u})P_{s}(x_{s})$ gives a
nonzero projector for finding particle $s$ at $x_{s}$ and particle
$u$ at $x'_{u}$. But for $u=s$, we have that $P_{s}(x'_{s})P_{s}(x_{s})=0$
for all $x'_{s}\neq x_{s}$. So the SN equations do not have a consistent
Born rule interpretation of the gravitational self-interaction term
for individual particles.

As further support for this conclusion, it is noteworthy that van
Wezel and van den Brink \cite{J.vanWezel2008} also concluded that
the SN equations cannot recover the Born rule for self-gravitating
particles, by studying the SN time-evolution of a two-state mass superposition.
Although they claim to be able to recover the Born rule for the two-state
superpositon by making the gravitational potential complex-valued,
they found that this approach does not work for superpositions of
more than two states. 

In spite of these arguments, it is still claimed by some researchers
\cite{Carlip2008,Salzman2005} that the standard Born rule interpretation
remains consistent for the SN equations because $|\psi|^{2}$ still
satisfies the familiar continuity equation,

\begin{equation}
\frac{\partial|\psi|^{2}}{\partial t}=\nabla\cdot\left[\frac{i\hbar}{2m}\left(\psi^{*}\overleftrightarrow{\nabla}\psi\right)\right].
\end{equation}
Adler's arguments are clearly correct, so how do we reconcile this
with (14)? It of course does not logically follow that if a density
such as $|\psi|^{2}$ satisfies a continuity equation like (14) then
it must necessarily be a probability density. If we interpret $m|\psi|^{2}$
as a real physical mass-matter density field in space-time, as opposed
to a mean mass density associated with an ensemble of particles, then
it can certainly satisfy (14) for all times without having a probability
interpretation. In this case, we would just interpret (14) as a statement
of local mass conservation for a massive `fluid' of density $m|\psi|^{2}$.
Note that this physical interpretation does not suffer from any inconsistency
problem when applied to gravitationally interacting and self-interacting
particles. For the $u\neq s$ case, we would say that (13) describes
the gravitational potential felt by `particle' $s$ at coordinate
$x_{s}$ from the matter density source $m_{u}|\psi_{u}(x'_{u},t)|^{2}$
for `particle' $u$. And for the $u=s$ case in (12), we would just
interpret this as the gravitational self-potential felt by the mass-matter
density field $m_{s}|\psi_{s}(x'_{s},t)|^{2}$ from the point $x'{}_{s}$
to the point $x{}_{s}$ within the density%
\footnote{This interpretation is in fact identical to the interpretation of
the gravitational self-potential for a continuous mass density in
classical Newtonian gravity theory. The only (albeit nontrivial) difference
in the SN case is that the mass density is defined in terms of the
mod-square of a wavefunction evolving by the SN equations. %
} 

We will therefore claim that the correct physical interpretation of
the SN equations is that they describe a world in which the wavefunction
in configuration space drives the dynamical evolution of a real physical
matter density field (or a set of N matter density fields in the N-particle
case) in 3-space, the evolving matter density field(s) sources an
evolving classical gravitational potential in 3-space, and this gravitational
potential couples back to the wavefunction, thereby altering the dynamical
evolution of the matter density field (the so-called gravitational
`back-reaction'). No statistical interpretation enters here. Of course,
this also means that the SN equations, correctly interpreted, have
no means of giving a consistent statistical description of the semiclassical
gravitational effects they predict, making them hard to take seriously
as the basis for an empirically viable semiclassical gravity theory. 

As a result of reaching the same conclusion about the empirical viability
of the SN equations, Adler poses the following questions (which he
does not attempt to answer) \cite{Adler2007}. Do the problems that
we have encountered (inconsistency with the Born rule interpretation)
indicate that a semiclassical approach to gravitation is inconsistent,
and hence that gravity must be quantized? Or do they only indicate
that a modifi{}cation of the Moeller\textendash{}Rosenfeld and SN
approach should be sought, which will make possible a consistent semiclassical
theory of gravitational effects? As we will later see, a modification
of the SN equations is indeed possible that leads to a physically
and statistically consistent nonrelativistic semiclassical theory
of gravitational effects.

\section{Existence of gravitational cat states}

Unfortunately, in addition to lacking a consistent statistical description
of gravitationally self-interacting particles, the correct physical
interpretation of the SN equations contains another problem for its
empirical viability - the existence of gravitational cat states for
macroscopic matter distributions. The SN equations can be shown to
admit cat states as follows. Elaborating the example by Ford \cite{Ford2005},
suppose we have a superposition state,

\begin{equation}
\psi_{cat}=\frac{1}{\sqrt{2}}\left[\phi_{left}+\phi_{right}\right],
\end{equation}
where each state in the superposition corresponds to a macroscopic
matter distribution in a distinct location (a 1000 kg mass occupying
a volume located on the left or right side of a room). Inserting $\psi_{cat}$
into (2) then gives

\begin{equation}
\nabla^{2}V_{cat}=4\pi G\left[\frac{m}{2}|\phi_{left}|^{2}+\frac{m}{2}|\phi_{right}|^{2}\right],
\end{equation}
with solution

\begin{equation}
V_{cat}=-G\int\frac{\frac{m}{2}\left[|\phi_{left}|^{2}+|\phi_{right}|^{2}\right]}{|x-x'|}d^{3}x'.
\end{equation}
In other words, we have a semiclassical gravitational fi{}eld, $\mathbf{g}_{cat}=-\nabla V_{cat}$,
which is an average of the fi{}elds due to the two distributions separately
(in this case, the gravitational field is the sum effect of two 500
kg masses on opposite sides of the room). By contrast, the simplistic
application of the measurement postulates of quantum mechanics would
tell us that if we were to measure the gravitational field with an
external test particle, $m_{test}$, the particle would feel a gravitational
force from a single 1000 kg matter density source occupying a single
location, and in different locations with equal frequency in multiple
measurement trials. As we saw in the previous section, such measurement
outcomes are not predicted by anything in the (correctly interpreted)
SN equations Moreover, Page and Geilker's torson balance pendulum
experiment has already disconfirmed the gravitational field predicted
by (17) for macroscopic superpositions \cite{Kiefer2012,D.N.Page1981}. 

It should be remarked that incorporating the effects of environmental
decoherence does not get rid of these cat states (for essentially
the same reason that decoherence doesn't solve the quantum measurement
problem); all decoherence can do is ensure that $\phi_{left}(q)\cdot\phi_{right}(q)\thickapprox0$
(i.e. $\phi_{left}$ and $\phi_{right}$ have disjoint supports in
configuration space) for all $q=(x_{1},...,x_{N})$ so that there
are no interference terms contributing to the right hand side of (16).
It should also be emphasized that the gravitational self-localization
effect discovered in numerical simulations by Salzman and Carlip \cite{Carlip2008},
Giulini and Grossardt \cite{DomenicoGiulini2011}, and Van Meter \cite{Meter2011},
does not solve the cat states problem either - all the self-localization
effect potentially does is ensure that each state in the superposition
will localize separate, 500 kg mass distributions around their respective
locations in 3-space. 

Thus we see that not only are these cat state solutions inconsistent
with experiment, there is no known physical mechanism based on the
SN equations alone that can suppress them. To incorporate a mechanism
that can requires modifying the SN equations in a non-trivial way,
as we will see in the next section.

\section{GRWmN}

Given the inconsistency with the Born rule intepretation and the existence
of gravitational cat states, it is clear that semiclassical gravity
based on the SN equations (or the Moeller-Rosenfeld equation) alone
is not empirically viable, nor really consistent with the postulates
of standard quantum mechanics. We suggest that one straightforward
way to modify the SN equations so as to give a consistent physical
and statistical description of gravitationally self-interacting particles
while also solving the cat states problem, is to develop a semiclassical
gravity theory based on an alternative collapse theory to standard
quantum mechanics, namely, the GRW theory \cite{G.C.Ghirardi1986};
and in particular, the variant of the GRW theory with matter density
ontology (GRWm) \cite{Allori2012}. We develop a nonrelativistic version
of such a theory in this section.

\subsection{Defining equations}

For a single-body system, the {}``GRWm-Newton'' (GRWmN) equations
are defined as follows. As in our interpretation of the SN equations,
we postulate the existence of a real physical matter density field
in space-time,

\begin{equation}
m(x,t)=m|\psi(x,t)|^{2},
\end{equation}
which we use as a source in the Newton-Poisson equation,

\begin{equation}
\nabla^{2}V(x,t)=4\pi Gm(x,t),
\end{equation}
where 

\begin{equation}
V(x,t)=-G\int\frac{m(x',t)}{|x-x'|}d^{3}x'.
\end{equation}
This gravitational `self-potential' couples back to the wavefunction
via the SN equation, 
\begin{equation}
i\hbar\partial_{t}\psi(x,t)=-\frac{\hbar^{2}}{2m}\nabla^{2}\psi(x,t)-Gm\int d^{3}x'\frac{m(x',t)}{|x-x'|}\psi(x,t),
\end{equation}
but now the wavefunction undergoes discrete and instantaneous intermittent
collapses according to the GRW collapse law. That is, the collapse
time $T$ occurs randomly with constant rate per system of $N\lambda=\lambda=10^{-16}s^{-1}$,
where the post-collapse wavefunction $\psi_{T+}=lim_{t\searrow T}\psi_{t}$
is obtained from the pre-collapse wavefunction $\psi_{T-}=lim_{t\nearrow T}\psi_{t}$
through multiplication by a Gaussian function,

\begin{equation}
\psi_{T+}(x)=\frac{1}{C}g(x-X)^{1/2}\psi_{T-}(x),
\end{equation}
where 

\begin{equation}
g(x)=\frac{1}{(2\pi\sigma^{2})^{3/2}}e^{-\frac{x^{2}}{2\sigma^{2}}}
\end{equation}
is the 3-D Gaussian function of width $\sigma=10^{-7}m$, and 

\begin{equation}
C=C(X)=\left(\int d^{3}xg(x-X)|\psi_{T-}(x)|^{2}\right)^{1/2}
\end{equation}
is the normalization factor. The collapse center $X$ is chosen randomly
with probability density $\rho(x)=C(x)^{2},$ and the space-time locations
of the collapses are given by the ordered pair $\left(X_{k,}T_{k}\right).$
Between collapses, the wavefunction just evolves by (21). The generalization
to an N-body system is as follows. We have N matter density fields
in 3-space as given by,

\begin{equation}
m(x,t)=\int dx'_{1}...dx'_{N}|\psi(x'_{1},...,x'_{N},t)|^{2}\sum_{i=1}^{N}m_{i}\delta^{3}(x-x'_{i}),
\end{equation}
which act as the source in the Newton-Poisson equation,
\begin{equation}
\nabla^{2}V(x,t)=4\pi G\int dx'_{1}...dx'_{N}|\psi(x_{1}'...x_{N}',t)|^{2}\underset{i=1}{\overset{N}{\sum}}m_{i}\delta^{3}(x-x'_{i}).
\end{equation}
The solution of (26) enters into the N-body SN equation,

\begin{equation}
i\hbar\partial_{t}\psi(x_{1}...x_{N},t)=-\underset{i=1}{\overset{N}{\sum}}\frac{\hbar^{2}}{2m_{i}}\nabla_{i}^{2}\psi(x_{1}...x_{N},t)-G\sum_{i,j=1}^{N}\int\frac{m_{i}m_{j}(x_{j}',t)}{|x_{i}-x_{j}'|}dx'_{1}...dx'_{N},
\end{equation}
and the solution of (27) undergoes collapse according to

\begin{equation}
\psi_{T+}(x_{1},...,x_{N})=\frac{1}{C}g(x_{i}-X)^{1/2}\psi_{T-}(x_{1},...,x_{N}),
\end{equation}
with probability density

\begin{equation}
\rho(X)=C(X)^{2}=\int dx'_{1}...dx'_{N}g(x'_{i}-X)|\psi_{T-}(x'_{1},...,x'_{N})|^{2},
\end{equation}
where $i$ is chosen randomly from $1,...,N.$ 

The equations of GRWmN for a single body say the following - a wavefunction
in 3-space, which evolves by (21) and undergoes the random collapse
process in (22), drives the dynamical evolution of a matter density
field in 3-space via (18). When the wavefunction collapses, it localizes
the matter density field around a randomly chosen point in 3-space,
with width of $10^{-7}$ meters, with the probability of the randomly
chosen point being largest where the mod-squared of the uncollapsed
wavefunction is largest, as indicated by (23). This evolving matter
density field also sources a gravitational potential in 3-space via
(20), and this potential couples back to the wavefunction via (21),
which in turn alters the evolution of the matter density field via
(20) again. 

For N-bodies, the wavefunction lives in configuration space $\mathbb{R}^{3N}$,
evolves by (27), and undergoes the collapse process in (28); this
wavefunction drives the dynamical evolution of N matter density fields
in 3-space via (25) so that when the wavefunction collapses, it randomly
localizes the matter density fields around randomly chosen (non-overlapping)
points in 3-space, each of width $10^{-7}$ meters, and with probability
density given by (29). As before, each of these matter density fields
acts as a source for a gravitational potential in 3-space that couples
back to the N-body wavefunction via (26)-(27), which in turn alters
the evolution of the matter density fields via (25) again.

\subsection{Consistent statistical description of gravitationally self-interacting
particles}

In contrast to the SN theory%
\footnote{By the {}``SN theory'', we refer to the (in our view) correct physical
interpretation of the SN equations discussed in section 2.%
}, GRWmN gives a consistent physical \textit{and} \textit{statistical}
description of gravitationally self-interacting particles. To see
this, let us reconsider the example in section 2. 

In GRWmN, $m_{u}|\psi_{u}(x'_{u},t)|^{2}$ is interpreted as the real
physical matter density field $m_{u}(x'_{u},t)$ for `particle'%
\footnote{We use parentheticals to emphasize that, just as in GRWm, there are
no particles in the fundamental physical ontology of GRWmN since the
matter density is a field distribution on space-time; particle-like
states can be said to emerge though when the wavefunction collapses
and localizes the matter density to a Gaussian distribution of very
narrow but finite width of $10^{-7}$meters. %
} $u$, with the probability for the collapse center $X_{u}$ given
by the separately defined GRW probability law, 

\begin{equation}
\rho(X{}_{u})=\int d^{3}x'_{u}g(x'_{u}-X_{u})|\psi_{T-}(x'_{u})|^{2},
\end{equation}
with collapse rate per particle of $\lambda=10^{-16}\frac{1}{s}$.
One would then say that the gravitational potential felt by particle
$s$ at coordinate $x_{s}$ from the matter density field $m_{u}(x'_{u},t)$
is

\begin{equation}
-\int d^{3}x'_{u}G\frac{m_{s}m_{u}(x'_{u},t)}{|x_{s}-x'_{u}|}
\end{equation}
and this is true whether the matter density for particle $u$ corresponds
to the uncollapsed wavefunction evolving by the SN equations or the
collapsed wavefunction defined by (22). Hence, if we now consider
the case $u=s$, we can see that GRWmN has no problem handling it
- as in our interpretation of the SN equations, we would just interpret
the expression,

\begin{equation}
-G\int d^{3}x'_{s}\frac{m_{s}m_{s}(x'_{s},t)}{|x_{s}-x'_{s}|},
\end{equation}
as the gravitational self-potential felt by the matter density $m_{s}(x'_{s},t)=m_{s}|\psi_{s}(x'_{s},t)|^{2}$
from the point $x'{}_{s}$ to the point $x{}_{s}$ within the density;
and this interpretation holds whether the matter density for particle
$s$ corresponds to the uncollapsed wavefunction or the collapsed
wavefunction. Moreover, it is clear that the probability for the matter
density for particle $s$ to collapse to center $X{}_{s}$ continues
to be consistently given by (30) for $u=s$.

\subsection{Suppression of gravitational cat states}

The GRWmN wavefunction, when evolving deterministically by the SN
equations, also admits gravitational cat states; but because the GRWmN
wavefunction undergoes random collapses according to (22) or (28),
which scales with the number of particles, those cat states are not
macroscopically observable. (Also, the gravitational field produced
by a cat state for a single elementary particle is presumably far
too weak to be experimentally measured.) For example, for a massive
object composed of Avogadro's number of particles, the collapse rate
is $\sim10^{7}\frac{1}{s}$. So the individual matter fields composing
the massive object will be localized around definite points in space
frequently enough to give the appearance of a macroscopic matter distribution
occupying a particular volume of space. 

Returning then to the example of a 1000 kg mass in the cat state $\psi_{cat}=\frac{1}{\sqrt{2}}\left[\phi_{left}+\phi_{right}\right]$,
it is clear that the number of systems needed in practice to compose
such a matter distribution would imply an astronomically faster collapse
rate; and when such collapses take place via (28), equations (26)
and (29) say that the result will be the appearance of a single 1000
kg mass localized on either the left or right side of the room (assuming
the collapse center $X$ for each system takes a binary outcome) with
equal frequency. More precisely, whereas the uncollapsed matter density
(using for simplicity the single-particle equation), 
\begin{equation}
m_{cat}(x,t)=m|\psi_{cat}|^{2}=\frac{1}{2}\left[m_{left}(x,t)+m_{right}(x,t)\right],
\end{equation}
produces the gravitational potential, 
\begin{equation}
V_{cat}=-G\int\frac{\frac{1}{2}\left[m_{left}(x',t)+m_{right}(x',t)\right]}{|x-x'|}d^{3}x',
\end{equation}
the GRW collapse gives either the state

\begin{equation}
\psi_{T+,left}(x)=\frac{1}{C_{left}}g(x-X_{left})^{1/2}\psi_{T-,cat}(x),
\end{equation}
or 

\begin{equation}
\psi_{T+,right}(x)=\frac{1}{C_{right}}g(x-X_{right})^{1/2}\psi_{T-,cat}(x),
\end{equation}
resulting in either the matter density 

\begin{equation}
m_{left}(x,t)=m|\psi_{T+,left}(x)|^{2},
\end{equation}
or 

\begin{equation}
m_{right}(x,t)=m|\psi_{T+,right}(x)|^{2},
\end{equation}
and either the gravitational potential

\begin{equation}
V_{left}=-G\int\frac{m_{left}(x',t)}{|x-x'|}d^{3}x',
\end{equation}
or

\begin{equation}
V_{right}=-G\int\frac{m_{right}(x',t)}{|x-x'|}d^{3}x'.
\end{equation}
The collapse center probability densities are correspondingly given
by

\begin{equation}
P_{left}(X_{left})=C_{left}^{2}=\frac{1}{2}\int d^{3}xg(x-X_{left})|\psi_{T-,cat}(x)|^{2},
\end{equation}
and

\begin{equation}
P_{right}(X_{right})=C_{right}^{2}=\frac{1}{2}\int d^{3}xg(x-X_{right})|\psi_{T-,cat}(x)|^{2}.
\end{equation}

This ensures that the gravitational field deflecting the test particle
in a single trial will look like it is due to only one matter density
source at only one of the locations, with the right amount of mass,
and with equal location frequency in N trials. In this way, the gravitational
field predicted by GRWmN is consistent with that observed in the Page
and Geilker experiment, in contrast to standard semiclassical gravity. 

Some comments are in order. First, it is known that the GRW collapse
causes a spontaneous increase in the average center-of-mass energy
for any system of N equal-mass particles \cite{G.C.Ghirardi1986,P.M.Pearle1994}.
So one would expect that the mass-equivalent of this increase in average
c.o.m. energy would contribute to the observed gravitational field
obtained from (39) or (40). In particular, it would slightly increase
the gravitational field strength produced by the collapsed matter
density on the left or right side and thus predict a slightly stronger
deflection of the test particle in the corresponding direction than
would be expected from just the bare mass value $m$. While this is
true, it is well-established \cite{G.C.Ghirardi1986,P.M.Pearle1994}
that the rate of increase in average c.o.m. energy due to the collapses
is so tiny that it is not macroscopically detectable with current
experimental capabilities%
\footnote{The rate of average energy increase is given in Ghirardi et al. \cite{G.C.Ghirardi1986}
by the formula $d\overline{E}/dt=(\hbar^{2}\lambda/4m\sigma^{2}).$
For the GRW values of $\lambda$ and $\sigma$, and using 1000 kg
as the mass, this gives a miniscule value of $\sim10^{-73}$ Joules
per second. It would then take $10^{63}$ years to increase the average
mass-energy by only $1$ mJ. %
}, making the contribution to the gravitational field produced by (39)
or (40) negligible as well.

Second, readers may notice a close parallel of the cat states problem
in semiclassical gravity with the quantum measurement problem in the
context of the Schroedinger-cat thought experiment. Indeed, it could
be said that the cat states problem is a manifestation of the measurement
problem in the context of semiclassical gravity, hence why the SN
theory (which is based on the standard quantum mechanics that originates
the measurement problem) suffers from it and GRWmN (which is free
of the measurement problem by virtue of the GRW collapse mechanism)
does not. Note, however, that not all proposed solutions to the measurement
problem extend successfully to semiclassical gravity. For example,
extending the Many-Worlds interpretations of Everett and Schroedinger
\cite{ValiaAllori2009} to the Moeller-Rosenfeld and SN equations
leads to the prediction that both branches of $\psi_{cat}$ should
continue to exist macroscopically and produce a semiclassical gravitational
field of the form in (17), which we already noted is ruled out by
the Page and Geilker experiment. This suggests that, to the extent
that one takes semiclassical gravity theory seriously, it can be used
as a testing ground for the robustness of any claimed solution to
the measurement problem in standard quantum mechanics%
\footnote{It is interesting to consider whether solutions to the measurement
problem based on non-local `hidden-variables', e.g. the de Broglie-Bohm
theory or Nelson's stochastic mechanics, can also solve the cat states
problem when extended to semiclassical gravity. This will be examined
in a forthcoming paper. %
}.

\subsection{Experimental prospects}

In addition to the GRW collapse process, the branches of the wavefunction
in GRWmN can undergo the gravitational self-localization effect observed
in numerical simulations of the SN equations for a free Gaussian wavepacket.
In particular, Giulini and Grossardt found that for $m=10^{10}$ amu
and initial width of 0.5 microns, a Gaussian wavepacket will undergo
self-localization, reach a minimum width of 0.4 microns in 30,000
seconds, and disperse again thereafter \cite{DomenicoGiulini2011}.
We would argue that GRWmN significantly improves the plausibility
of this prediction of the SN equations, considering that GRWmN gives
a physically and statistically consistent semiclassical description
of gravitational effects and the SN theory does not. 

As it has been suggested \cite{DomenicoGiulini2011,K.Hornberger2012,P.Asenbaum(2013)}
that molecular interferometry experiments with macromolecule clusters
may eventually reach this mass scale, it is natural to ask if GRW
collapse might also be observable at this mass scale and perhaps happen
`on top of' the self-localization effect. If we make the generous
assumption that in GRWmN a mass of $10^{10}$ amu corresponds to $10^{10}$
particles of 1 amu, this gives an approximate collapse rate of $10^{-6}\frac{1}{s}$,
or $10^{6}s$ for each collapse. In other words, to have a chance
of observing a single GRW collapse event, we would have to maintain
the coherence time of the wavepacket for a minimum of about 33 times
longer than the timescale it takes the self-localization to reach
the minimum width. 

It remains to be seen whether technological advancements in molecular
interferometry that allow for maintaining coherence times of 30,000
seconds will also allow for maintaining coherence times of $10^{6}s$
or greater. Even so, we note that if self-localization is not observed
at the mass scale predicted by the SN equations used in GRWmN, this
will be sufficient to falsify GRWmN as a semiclassical theory of gravity.
And if self-localization is observed, it would be strong evidence
for GRWmN or some dynamical collapse variant of GRWmN%
\footnote{It should also be made clear to what extent experimental confirmation
or falsification of the predictions of GRWmN could also be taken as
experimental confirmation or falsification of the GRW approach to
quantum theory more generally. While self-localization not being observed
at the predicted mass scale would falsify GRWmN, it would not be sufficient
to falsify the GRW approach more generally - if gravity is fundamentally
quantized, it is possible that the correct description of quantum
gravity is not the Wheeler-deWitt equation of canonical quantum gravity
(from which the SN equations are obtained as the semiclassical approximation),
but a different approach that has a semiclassical approximation that
does not give the SN equations (string theory would be one example);
or, if gravity is fundamentally emergent, it is possible that the
correct approach to emergent gravity is something different than the
approaches which suggest the semiclassical Einstein equation as the
emergent description of gravity (the induced-gravity approach of Sakharov
\cite{Visser2002} would be such an example). It is then logically
possible that the GRW theory could work in these alternative approaches.
On the other hand, the observation of self-localization would, as
we said, be strong, indirect evidence for the existence of the GRW
collapse dynamics (or a similar collapse dynamics such as the CSL
dynamics \cite{Pearle}), and thus GRWmN. And if, on top of the self-localization
effect, the GRW collapse effect is observed directly, or if statistical
consequences of the GRW collapse process are confirmed directly, this
would be direct confirmation of GRWmN and the GRW dynamics it employs. %
}.

It is also worth commenting on Yang et al.'s \cite{H.Yang2013} recent
observation that the SN equations predict that a single macroscopic
quantum object (modeled by a squeezed Gaussian state) evolving in
a harmonic potential has a quantum uncertainty that evolves at a different
frequency than the standard quantum mechanical eigenfrequency, and
that testing such a prediction is within the capabilities of state-of-the-art
optomechanical experiments. Yang et al. compute quantum expectation
values of quantum operators and implicitly assume the applicability
of the Born rule and quantum projection postulate. But given our observations
about the inconsistency of the Born rule as applied to the SN equations,
this should raise doubts about the plausibility of the method by which
they obtained their prediction. Thus, it would seem important to analyze
the physical system they consider in the context of a semiclassical
gravity theory that does have a consistent statistical description,
e.g. GRWmN. In fact, we would expect GRWmN to give expectation values
very close to that found by Yang et al., given that the GRW probability
density formula (29), and expectation values of GRW operators computed
with it, in general gives values very close to - but not exactly equal
to - the Born rule distribution of quantum mechanics and expectation
values of quantum operators computed with it \cite{G.C.Ghirardi1986,S.Goldstein2012}.
If correct, this would give us much greater confidence in Yang et
al.'s prediction of a frequency difference, and it would be yet another
route to experimentally testing GRWmN. Perhaps an even more feasible
prediction to test than either the self-localization effect or the
GRW collapse effect.

\section{Discussion}

To summarize, we reviewed the problem of the inconsistency of the
SN equations with the Born rule interpretation and the problem of
gravitational cat states. We then presented a GRW modification of
Newtonian semiclassical gravity, called GRWmN, to solve these problems.
In this way, we showed that one can have a nonrelativistic semiclassical
gravity theory related to the Moeller-Rosenfeld equation that's empirically
viable in its physical and statistical description of gravitationally
self-interacting particles and macroscopic mass superpositions. In
addition, we indicated how GRWmN could in principle be experimentally
testable with future molecular interferometry experiments, and how
it can be used to revise the predictions of a recently proposed macroscopic
test of Newtonian semiclassical gravity based on the SN equations. 

To the extent that GRWmN is based on a version of GRW theory with
a `primitive ontology'%
\footnote{Primitive ontology is defined in \cite{Allori2012} as just the {}``variables
describing the distribution of matter in 4-dimensional space-time''. %
} (GRWm), one could ask if other GRW theories with primitive ontologies
could be extended to semiclassical gravity. Along with GRWm, perhaps
the best known GRW theory with a primitive ontology is GRWf \cite{Allori2012},
where f stands for the `flash' ontology, i.e. an ontology in which
matter is represented by the space-time locations of the collapsed
wavefunction, or the set $F=\left\{ (X_{1},T_{1}),...,(X_{k},T_{k}),...\right\} $.
Obviously the flashes can't, on their own, be used as a source for
a semiclassical gravitational field, as they are just space-time locations.
One could perhaps modify GRWf so that at each flash a point mass is
spontaneously produced, which then acts as a source for a gravitational
potential%
\footnote{This possibility was suggested by T. Norsen {[}personal communication{]}.%
}. However, such a theory would entail gross violation of mass conservation
every time a flash is produced, making it seem rather contrived, even
though logically possible. A more straightforward and natural way
we can see to retain the flashes in GRWf and also extend it to a Newtonian
theory of semiclassical gravity is to add the presence of a matter
density field in space-time which, upon collapse, is localized around
the flashes. But this is essentially what happens in GRWm, to the
extent that the GRW collapse localizes the matter density field around
the space-time locations of the collapsed wavefunction. It would then
seem artificial and contrived to continue to insist that the flashes
constitute the primitive ontology in this approach. So the more natural
GRW approach to semiclassical gravity seems to suggest the matter
density ontology as the more natural one to describe the physical
world, even for cases when gravitational self-interaction effects
can be neglected.

Concerning the collapse mechanism, although we based our formulation
of semiclassical gravity on the discrete and instantaneous GRW collapse,
it seems entirely possible to also use a continuous and non-instantaneous
collapse mechanism such as CSL (Continuous Spontaneous Localization)
\cite{Pearle}, together with a matter density field ontology (see
for example \cite{Diosi2012A}). We used the GRW collapse mechanism
because it is mathematically the simplest dynamical collapse process
for formulating a Newtonian semiclassical gravity model, and because
it is already used in GRWm, a theory which is known to match the predictions
of standard nonrelativistic quantum mechanics for all current experiments
while also having a primitive ontology that can be naturally extended
to semiclassical gravity. (Note that Weinberg's recently proposed
collapse model \cite{Weinberg2012} contains the GRW collapse process
as a special case, and could certainly be used as well.) On the other
hand, using the CSL mechanism would have the advantage that it would
be testable with molecular interferometry at much lower mass scales
(between $10^{6}$ and $10^{8}$ amu, according to Nimmrichter et
al. \cite{S.Nimmrichter2011}) than GRWmN, given that the collapse
rate in CSL scales quadratically with the total mass $m$ of a system.

Although we restricted our theory to Newtonian semiclassical gravity,
it certainly seems possible to extend it to the semiclassical Einstein
equation for both first and second quantized wavefunctions. As we
will show in a forthcoming paper, what needs to change for this extension
are 1) the collapse law and the matter density field definition -
rather than using the nonrelativistic GRW collapse law, we would instead
use the relativistic GRW collapse law developed by Tumulka in \cite{Tumulka2006},
and the relativistic stress-energy-momentum tensor field proposed
by Bedingham et al. in \cite{D.Bedingham2012}; 2) for first quantized
wavefunctions, the evolution equations should be the first quantized
Klein-Gordon or Dirac equation coupled to the semiclassical Einstein
equation \cite{D.Giulini2012}, while for second quantized wavefunctions
one should use the functional Schroedinger or functional Dirac equation
coupled to the semiclassical Einstein equation \cite{Reis1999}. It
should be noted however that such a theory will have an inconsistency
that also plagues standard semiclassical gravity - the covariant divergence
of the right hand side of the semiclassical Einstein equation is nonzero
upon wavefunction collapse, while the covariant divergence of the
left hand side is always zero (i.e. the Bianchi identity). As Tumulka
has noted {[}personal communication{]}, this inconsistency may mean
that the semiclassical Einstein equation with a wavefunction undergoing
the GRW evolution does not possess any solutions. This is a question
that needs further research%
\footnote{Wald \cite{Wald1994} has developed a prescription for measurement
in standard semiclassical gravity that can be given a collapse interpretation
and also satisfies $<T_{nm}>;_{m}=0$, but it is not clear to us if
this prescription can be extended to a GRW theory. %
}, in which case, a GRW version of semiclassical gravity would seem
to fare no better here at the moment than standard semiclassical gravity. 

A statistical inconsistency in semiclassical gravity theories such
as the SN equations has also been claimed by Salcedo \cite{Salcedo2012},
and which at first glance might seem to apply to GRWmN and its possible
general relativistic extension. Apart from the fact that Salcedo assumes
the applicability of the Born rule interpretation and quantum project
postulate, which as we saw have no consistent application to the SN
equations, Barcelo et al. \cite{C.Barcelo2012} have noted that one
can interpret the violation of Salcedo's statistical consistency criterion
as an indication that semiclassical gravity theories need to be complemented
with the selection of a natural basis in Hilbert space - a notion
that seems compatible with the notion of a {}``pointer basis'' in
quantum decoherence theory (which can readily be incorporated with
the GRW formalism \cite{Schlosshauer2004}), as well as the fact that
the GRW collapse mechanism selects out the pointer basis as the universally
preferred basis \cite{G.C.Ghirardi1986,Schlosshauer2004}. 

One might also be concerned with Gisin's proof \cite{Gisin1989} that
any nonlinear deterministic wave equation, such as the SN equations
or the semiclassical Einstein equation, would allow for superluminal
signaling. However, several researchers \cite{Polchinski1991,Jordan1993,Czachor1996,Salzman2005}
have identified loopholes in Gisin's proof that arguably would invalidate
the application of the proof to the SN equations. Moreover, there
is a more direct objection to the claim that Gisin's proof applies
to the SN equations, let alone GRWmN - the proof assumes the validity
of the Born rule and quantum projection postulate, which we have seen
are inconsistent with the SN equations, and of course are not part
of the GRWmN theory. One might then ask whether the GRW collapse process
allows for superluminal signaling. Fortunately, Gisin also showed
in \cite{Gisin1989} that a nonlinear stochastic collapse modification
of quantum mechanics equivalent to the GRW theory does not suffer
from the problem of superluminal signaling. Hence GRWmN, insofar as
it uses the GRW collapse process, will not allow for superluminal
signaling either. 

Along with these (real and alleged) inconsistencies, a general relativistic
extension of GRWmN would also inherit the stability problem of standard
semiclassical gravity (i.e. the fact that the semiclassical Einstein
equation is a fourth-order system means that some solutions have runaway
behavior), and the formally divergent expectation value of $T_{nm}$
\cite{Ford2005}. But these latter problems seem to be manageable
- Ford \cite{Ford2005} has noted for standard semiclassical gravity
that there exist adequate renormalization procedures for $<T_{nm}>$,
and there exist reasonable proposals for solving the stability problem
by either reformulating the semiclassical Einstein equation as an
integro-differential equation to eliminate the runaway solutions or
by regarding the semiclassical gravity theory as valid only for spacetimes
that pass a certain stability criterion.

In light of all these difficulties that seem to arise in attempting
to formulate a consistent theory of semiclassical gravity, we would
like to anticipate a potential critic who might ask why one should
care about doing so. Especially since many physicists ultimately want
a full quantum theory of gravity which presumably won't have the consistency
problems of semiclassical gravity theory. 

First, as mentioned in the introduction, one of the major approaches
to quantum gravity, canonical quantum gravity, has the semiclassical
Einstein equation as a prediction of its semiclassical approximation
\cite{Kiefer2012,B.L.Hu2008}. So the semiclassical limit of canonical
quantum gravity - on which some of the calculational results of canonical
quantum gravity are based \cite{Kiefer2012} - inherits all the problems
we've discussed as semiclassical gravity theory taken as fundamental.
This would seem to suggest that canonical quantum gravity would also
benefit from a GRW-type modification%
\footnote{In this connection, it would be interesting to explore whether a GRW
version of canonical quantum gravity with matter density ontology
would, in its semiclassical Newtonian approximation, reduce to our
GRWmN theory. %
}. 

Second, it has been argued by some researchers that gravity may be
an emergent, collective phenomenon, in which case it would be misguided
to try and quantize it \cite{Volovik2013,S.Finazzi2012,Hu2009,C.Barcelo2001,Visser2002,Hu2011,Jacobson1995,Boughn2009}.
This view motivates models of emergent gravity mentioned earlier,
which have the semiclassical Einstein equation as the emergent description
of the coupling between quantum theory and gravity \cite{Volovik2013,S.Finazzi2012,Hu2009}.
Like with canonical quantum gravity, then, emergent gravity approaches
could also benefit from a GRW-type modification.

Third, we reiterate that semiclassical gravity (in its standard formulation),
in spite of its difficulties, has been used to derive several key
results in theoretical astrophysics - Hawking radiation, cosmological
perturbations in cosmic inflation, particle pair production in expanding
spacetimes, the creation of naked black hole singularities, traversable
wormhole solutions, warp drive spacetimes, as just a few examples
\cite{Ford2005,N.D.Birrell1982}. There are also the more recent modifications
of semiclassical Einstein gravity mentioned earlier, such as stochastic
gravity and gravitational aether-theory, which offer new physical
insights into the aforementioned astrophysical phenomena. To the best
of our understanding, both stochastic gravity and gravitational aether-theory
share all the previously discussed technical difficulties of standard
semiclassical gravity, and so would presumably also benefit from a
GRW-type modification%
\footnote{In regards to stochastic gravity, it should be noted that Hu and Verdaguer
\cite{B.L.Hu2008} developed it to deal with cat states (or large
\textquotedblleft{}stress tensor fluctuations\textquotedblright{}
as they put it). They propose an \textquotedblleft{}Einstein-Langevin\textquotedblright{}
equation, a linear equation for a stochastic metric perturbation $h_{nm}$
that takes into account the fluctuations of the stress tensor and
is superposed on top of the metric $g_{nm}$ satisfying the semiclassical
Einstein equation. It seems to us, however, that the Einstein-Langevin
equation they propose would \emph{prima facie} still suffer from the
cat states problem for macroscopic superpositions since $g_{nm}$
is still a solution of the semiclassical Einstein equation, and since
the solution of the Einstein-Langevin equation still depends on $<\psi|\hat{T}_{nm}|\psi>$.
We would therefore expect that a test particle sent to probe the stochastic
gravitational field $g_{nm}+h_{nm}$ for a macroscopic superposition
of two masses will not get deflected by a definite mass distribution
at a definite spacetime location, but rather will feel a stochastically
fluctuating gravitational field that will be an average of the fluctuating
gravitational fields due to two spatially separate masses. An analysis
of this situation in the context of the Newtonian limit of stochastic
gravity has yet to be done, to the best of our knowledge. However,
we would expect the Newtonian case to give the same conclusion unless
the stochasticity of the metric perturbations can induce dynamical
collapse in the pointer basis, as is the case with the phenomenological
stochastic noise field in CSL \cite{Pearle}.%
}. 

Fourth, unlike the quantum gravitational phenomena predicted by most
quantum gravity theories, many of the predictions of semiclassical
gravity theories may be empirically testable in the near future. On
the observational astrophysics side, the density perturbations in
the CMB spectrum predicted by semiclassical gravitational effects
in eternal cosmic inflation \cite{B.L.Hu2008} may soon be tested
with the Planck Satellite's mapping of the CMB power spectrum \cite{StephenM.Feeney2011}.
Aslanbeigi et al. \cite{S.Aslanbeigi2011} have also shown that gravitational
aether-theory makes observationally testable predictions for the gravitational
constant of radiation vs. nonrelativistic matter, as well as for the
intrinsic gravitomagnetic effect. On the experimental condensed matter
side, it has been proposed by Weinfurtner et al. \cite{SilkeWeinfurtner2011}
that a condensed matter analogue of Hawking radiation may be observed
in experiments using superfluids with supersonic flow velocities.
It has also been shown by Barcelo et al. \cite{CarlosBarcelo2003}
that the prediction of pair production on expanding spacetimes could
be tested in a BEC that simulates a quantum field evolving on an expanding
spacetime. And of course, on the experimental atomic/molecular physics
side, we have the prediction discussed earlier of gravitational self-localization
from the SN equations, which may be testable in molecular interferometry
experiments of the near future, as well as the quantum-uncertainty
frequency difference for a macroscopic object in a harmonic potential,
which should be testable with state-of-the-art optomechanical experiments.
Given the potential testability of the various astrophysical predictions
of standard semiclassical gravity, and the problematic physical foundations
on which standard semiclassical gravity rests, it also seems worthwhile
to re-analyze these predictions with a GRW approach to see if the
GRW approach gives differing predictions that could be observed in
said tests. In this regard, it has recently come to our attention
{[}Struyve, personal communication{]} that Landau et al. \cite{SusanaJ.Landau2012}
have constructed and applied CSL variants of GRWmN to a scalar quantum
field undergoing cosmic inflation, and shown that such models leads
to different CMB power spectrums than the standard, scale-invariant
Harrison-Zel'dovich one. They have also shown that such collapse models
are currently consistent with data on the CMB from WMAP. It remains
to be seen if such deviations are also consistent with data on the
CMB from the more recent Planck satellite. 

In sum, we believe semiclassical gravity is a very worthwhile approach
for trying to consistently incorporate quantum mechanics and gravity.
And we believe dynamical collapse versions of semiclassical gravity,
such as GRWmN and its possible relativistic extensions, provide a
much more promising route to a consistent formulation of semiclassical
gravity than does standard quantum mechanics.

\section{Acknowledgments}

I thank Roderich Tumulka and Sheldon Goldstein for clarifying discussions
and helpful comments on an earlier draft of this paper. I also thank
Bei-Lok Hu for useful comments on semiclassical gravity theories,
as well as two anonymous referees, one for suggesting two useful references
and another for asking a simple question that resulted in significant
improvements to the paper. 

\bibliographystyle{unsrt}
\bibliography{GRWm_refs}

\begin{thebibliography}{10}

\bibitem{Moeller1962}
C.~Moeller.
\newblock {\em Les Theories Relativistes de la Gravitation}.
\newblock Colloques Internationaux CNRS 91, CNRS 91, Paris, 1962.

\bibitem{Rosenfeld1963}
L.~Rosenfeld.
\newblock On the quantization of fields.
\newblock {\em Nucl. Phys.}, 40:353, 1963.

\bibitem{Carlip2008}
S.~Carlip.
\newblock Is quantum gravity necessary?
\newblock {\em Class. Quant. Grav.}, 25:154010, 2008,
  http://arxiv.org/abs/0803.3456.

\bibitem{Kiefer2012}
C.~Kiefer.
\newblock {\em Quantum Gravity, third edition}.
\newblock (Oxford University Press, Oxford), 2012.

\bibitem{B.L.Hu2008}
B.~L. Hu and E.~Verdaguer.
\newblock Stochastic gravity: Theory and applications.
\newblock {\em Living Rev. Relativity}, 11:3, 2008,
  http://arxiv.org/abs/0802.0658.

\bibitem{Volovik2013}
G.~E. Volovik.
\newblock {\em Novel Superfluids}, chapter The Superfluid Universe, pages
  570--618.
\newblock International Series of Monographs on Physics 156, Volume 1, 2013,
  http://arxiv.org/abs/1004.0597.

\bibitem{S.Finazzi2012}
S.~Finazzi, S.~Liberati, and L.~Sindoni.
\newblock Cosmological constant: a lesson from bose-einstein condensates.
\newblock {\em Phys. Rev. Lett.}, 108:071101, 2012,
  http://arxiv.org/abs/1103.4841.

\bibitem{Hu2009}
B.~L. Hu.
\newblock Emergent/quantum gravity: Macro/micro structures of spacetime.
\newblock {\em J. Phys. Conf. Ser.}, 174:012015, 2009,
  http://arxiv.org/abs/0903.0878.

\bibitem{C.Barcelo2001}
C.~Barcelo, M.~Visser, and S.~Liberati.
\newblock Einstein gravity as an emergent phenomenon?
\newblock {\em Int. J. Mod. Phys.}, D10:799--806, 2001,
  http://arxiv.org/abs/gr-qc/0106002.

\bibitem{Ford2005}
L.~H. Ford.
\newblock {\em 100 Years of Relativity - Space-time Structure: Einstein and
  Beyond}, chapter Spacetime in Semiclassical Gravity.
\newblock World Scientific, 2005, http://arxiv.org/abs/grqc/0504096.

\bibitem{N.D.Birrell1982}
N.~D. Birrell and P.~C.~W. Davies.
\newblock {\em Quantum Fields in Curved Space}.
\newblock Cambridge University Press, 1982.

\bibitem{S.Aslanbeigi2011}
S.~Aslanbeigi, G.~Robbers, B.~Z. Foster, K.~Kohri, and N.~Afshordi.
\newblock Phenomenology of gravitational aether as a solution to the old
  cosmological constant problem.
\newblock {\em Phys. Rev. D}, 84:103522, 2011, http://arxiv.org/abs/1106.3955.

\bibitem{M.Saravani2012}
M.~Saravani, N.~Afshordi, and R.~B.~Mann (2012).
\newblock Empty black holes, firewalls, and the origin of bekenstein-hawking
  entropy.
\newblock http://arxiv.org/abs/1212.4176.

\bibitem{Salzman2005}
P.~J. Salzman.
\newblock {\em Investigation of the time dependent Schroedinger-Newton
  equation}.
\newblock PhD thesis, University of California - Davis, 2005.

\bibitem{Penrose1996}
R.~Penrose.
\newblock On gravity's role in quantum state reduction.
\newblock {\em General Relativity and Gravitation}, 28:581--600, 1996.

\bibitem{Penrose1998}
R.~Penrose.
\newblock Quantum computation, entanglement and state reduction.
\newblock {\em Phil. Trans. R. Soc. Lond. A}, 356:1927--1939, 1998,
  http://citeseerx.ist.psu.edu/viewdoc/download?doi=10.1.1.84.7047\&rep=rep1\&type=pdf.

\bibitem{Diosi1984}
L.~Diosi.
\newblock Gravitation and quantum-mechanical localization of macro-objects.
\newblock {\em Phys. Lett. A}, 105:199--202, 1984.

\bibitem{D.Giulini2013}
D.~Giulini and A.~Grossardt.
\newblock Gravitationally induced inhibitions of dispersion according to a
  modified schroedinger-newton equation for a homogeneous-sphere potential.
\newblock {\em Class. Quant. Grav.}, 30:155018, 2013,
  http://arxiv.org/abs/1212.5146.

\bibitem{Meter2011}
J.~R. van Meter.
\newblock Schroedinger-newton "collapse" of the wave function.
\newblock {\em Class. Quant. Grav.}, 28:215013, 2011,
  http://arxiv.org/abs/1105.1579.

\bibitem{H.Yang2013}
H.~Yang, H.~Miao, D-S Lee, B.~Helou, and Y.~Chen.
\newblock Macroscopic quantum mechanics in a classical spacetime.
\newblock {\em Phys. Rev. Lett.}, 110:170401, 2013,
  http://arxiv.org/abs/1210.0457.

\bibitem{DomenicoGiulini2011}
D.~Giulini and A.~Grossardt.
\newblock Gravitationally induced inhibitions of dispersion according to the
  schroedinger-newton equation.
\newblock {\em Class. Quantum Grav.}, 28:195026, 2011,
  http://arxiv.org/abs/1105.1921.

\bibitem{Diosi2012A}
L.~Diosi.
\newblock Gravity-related wave function collapse: mass density resolution.
\newblock In {\em Invited talk at Sixth International Workshop DICE2012}, 2012,
  http://arxiv.org/abs/1302.5365.

\bibitem{Diosi2012B}
L.~Diosi.
\newblock Gravity-related wave function collapse: Is superfluid he exceptional?
\newblock In {\em Invited talk at the international workshop "Horizon of
  quantum physics: from foundations to quantum enabled technologies".}, 2012,
  http://arxiv.org/abs/1302.5364.

\bibitem{Adler2007}
S.~L. Adler.
\newblock Comments on proposed gravitational modifications of schroedinger
  dynamics and their experimental implications.
\newblock {\em J. Phys. A.}, 40:755--764, 2007,
  http://arxiv.org/abs/quant-ph/0610255.

\bibitem{K.Hornberger2012}
K.~Hornberger, S.~Gerlich, P.~Haslinger, S.~Nimmrichter, and M.~Arndt.
\newblock Colloquium: Quantum interference of clusters and molecules.
\newblock {\em Rev. Mod. Phys.}, 84:157, 2012, http://arxiv.org/abs/1109.5937.

\bibitem{P.Asenbaum(2013)}
P.~Asenbaum, S.~Kuhn, S.~Nimmrichter, U.~Sezer, and M.~Arndt (2013).
\newblock Cavity cooling of free silicon nanoparticles in high-vacuum.
\newblock http://arxiv.org/abs/1306.4617.

\bibitem{W.Marshall2003}
W.~Marshall, C.~Simon, R.~Penrose, and D.~Bouwmeester.
\newblock Towards quantum superpositions of a mirror.
\newblock {\em Phys. Rev. Lett.}, 91:130401, 2003,
  http://arxiv.org/abs/quant-ph/0210001.

\bibitem{D.N.Page1981}
D.~N. Page and C.~D. Geilker.
\newblock Indirect evidence of quantum gravity.
\newblock {\em Phys. Rev. Lett.}, 47:979--982, 1981.

\bibitem{G.C.Ghirardi1986}
G.~C. Ghirardi, A.~Rimini, and T.~Weber.
\newblock Unified dynamics for microscopic and macroscopic systems.
\newblock {\em Phys. Rev. D.}, 34::470--491, 1986.

\bibitem{Allori2012}
V.~Allori, S.~Goldstein, R.~Tumulka, and N.~Zanghi (2012).
\newblock Predictions and primitive ontology in quantum foundations: a study of
  examples.
\newblock http://arxiv.org/abs/1206.0019.

\bibitem{D.Giulini2012}
D.~Giulini and A.~Grossardt.
\newblock The schroedinger-newton equation as non-relativistic limit of
  self-gravitating klein-gordon and dirac fields.
\newblock {\em Class. Quant. Grav.}, 29:215010, 2012,
  http://arxiv.org/abs/1206.4250.

\bibitem{F.SiddharthaGuzman2003}
F.~Siddhartha Guzman and L.~Arturo Urena-Lopez.
\newblock Newtonian collapse of scalar field dark matter.
\newblock {\em Phys. Rev. D}, 68:024023, 2003,
  http://arxiv.org/abs/astro-ph/0303440.

\bibitem{T.W.Kibble1980}
T.W. Kibble and S.~Randjbar-Daemi.
\newblock Non-linear coupling of quantum theory and classical gravity.
\newblock {\em J. Phys. A.}, 13:141, 1980.

\bibitem{Boughn2009}
S.~Boughn.
\newblock Nonquantum gravity.
\newblock {\em Found. Phys.}, 39:331--351, 2009,
  http://arxiv.org/abs/0809.4218.

\bibitem{J.vanWezel2008}
J.~van Wezel and J.~van~den Brink.
\newblock Schroedinger-newton equation as a possible generator of quantum state
  reduction.
\newblock {\em Phil. Mag.}, 88:1659--1671, 2008,
  http://arxiv.org/abs/0803.4488.

\bibitem{P.M.Pearle1994}
P.~M. Pearle and E.~Squires.
\newblock Bound state excitation, nucleon decay experiments, and models of wave
  function collapse.
\newblock {\em Phys. Rev. Lett.}, 73:1--5, 1994,
  http://cds.cern.ch/record/260011/files/P00021809.pdf.

\bibitem{ValiaAllori2009}
Valia Allori, Sheldon Goldstein, Roderich Tumulka, and Nino Zanghi.
\newblock Many-worlds and schroedinger's first quantum theory.
\newblock 2009, http://arxiv.org/abs/0903.2211.

\bibitem{Visser2002}
M.~Visser.
\newblock Sakharov's induced gravity: a modern perspective.
\newblock {\em Mod. Phys. Lett.}, A17:977--992, 2002,
  http://arxiv.org/abs/gr-qc/0204062.

\bibitem{Pearle}
P.~Pearle (2012).
\newblock Collapse miscellany.
\newblock To be included in a volume honoring the 80th birthday of Yakir
  Aharanov. http://arxiv.org/abs/1209.5082.

\bibitem{S.Goldstein2012}
S.~Goldstein, R.~Tumulka, and N.~Zanghi.
\newblock The quantum formalism and the grw formalism.
\newblock {\em Journal of Statistical Physics}, 149:142--201, 2012,
  http://arxiv.org/abs/0710.0885.

\bibitem{Weinberg2012}
S.~Weinberg.
\newblock Collapse of the state vector.
\newblock {\em Phys. Rev. A}, 85:062116, 2012, http://arxiv.org/abs/1109.6462.

\bibitem{S.Nimmrichter2011}
S.~Nimmrichter, K.~Hornberger, P.~Haslinger, and M.~Arndt.
\newblock Testing spontaneous localization theories with matter-wave
  interferometry.
\newblock {\em Phys. Rev. A}, 83:043621, 2011, http://arxiv.org/abs/1103.1236.

\bibitem{Tumulka2006}
R.~Tumulka.
\newblock A relativistic version of the ghirardi-rimini-weber theory.
\newblock {\em J. Statist. Phys.}, 125:821--840, 2006,
  http://arxiv.org/abs/quant-ph/0406094.

\bibitem{D.Bedingham2012}
D.~Bedingham, D.~Duerr, G.~Ghirardi, S.~Goldstein, R.~Tumulka, and N.~Zanghi
  (2012).
\newblock Matter density and relativistic theories of wave function collapse.
\newblock http://arxiv.org/abs/1111.1425.

\bibitem{Reis1999}
H.~C. Reis.
\newblock Quantum evolution of inhomogeneities in curved space.
\newblock {\em Int. J. Mod. Phys.}, A14:1633--1650, 1999,
  http://arxiv.org/abs/hep-th/0108175.

\bibitem{Wald1994}
R.~M. Wald.
\newblock {\em Quantum Field Theory in Curved Spacetime and Black Hole
  Thermodynamics.}
\newblock Chicago University Press, Chicago., 1994.

\bibitem{Salcedo2012}
L.~L. Salcedo.
\newblock Statistical consistency of quantum-classical hybrids.
\newblock {\em Phys. Rev. A}, 86:022127, 2012, http://arxiv.org/abs/1201.4237.

\bibitem{C.Barcelo2012}
C.~Barcelo, R.~Carballo-Rubio, L.~J. Garay, and R.~Gomez-Escalante.
\newblock Hybrid classical-quantum formulations ask for hybrid notions.
\newblock {\em Phys. Rev. A}, 86:042120, 2012, http://arxiv.org/abs/1206.7036.

\bibitem{Schlosshauer2004}
M.~Schlosshauer.
\newblock Decoherence, the measurement problem, and interpretations of quantum
  mechanics.
\newblock {\em Rev. Mod. Phys.}, 76:1267--1305, 2004,
  http://arxiv.org/abs/quant-ph/0312059.

\bibitem{Gisin1989}
N.~Gisin.
\newblock Stochastic quantum dynamics and relativity.
\newblock {\em Helvetica Physica Acta}, 62:363--371, 1989.

\bibitem{Polchinski1991}
J.~Polchinski.
\newblock Weinberg's nonlinear quantum mechanics and the
  einstein-podolsky-rosen paradox.
\newblock {\em Phys. Rev. Lett.}, 66:397--400, 1991.

\bibitem{Jordan1993}
T.~F. Jordan.
\newblock Reconstructing a nonlinear dynamical framework for testing quantum
  mechanics.
\newblock {\em Annals of Physics}, 225:83--113, 1993.

\bibitem{Czachor1996}
M.~Czachor.
\newblock Nonlinear schroedinger equation and two-level atoms.
\newblock {\em Phys. Rev. A}, 53:1310--1315, 1996,
  http://arxiv.org/abs/quant-ph/9501007.

\bibitem{Hu2011}
B.~L. Hu.
\newblock Gravity and nonequilibrium thermodynamics of classical matter.
\newblock {\em Int. J. Mod. Phys.}, D20:697--716, 2011,
  http://arxiv.org/abs/1010.5837.

\bibitem{Jacobson1995}
T.~Jacobson.
\newblock Thermodynamics of spacetime: The einstein equation of state.
\newblock {\em Phys. Rev. Lett.}, 75:1260--1263, 1995,
  http://arxiv.org/abs/gr-qc/9504004.

\bibitem{StephenM.Feeney2011}
S.~M. Feeney, M.~C. Johnson, D.~J. Mortlock, and H.~V. Peiris.
\newblock First observational tests of eternal inflation: analysis methods and
  wmap 7-year results.
\newblock {\em Phys. Rev. D.}, 84:043507, 2011, http://arxiv.org/abs/1012.3667.

\bibitem{SilkeWeinfurtner2011}
S.~Weinfurtner, E.~W. Tedford, M.~C.~J. Penrice, W.~G. Unruh, and G.~A.
  Lawrence.
\newblock Measurement of stimulated hawking emission in an analogue system.
\newblock {\em Phys. Rev. Lett.}, 106:021302, 2011,
  http://arxiv.org/abs/1008.1911.

\bibitem{CarlosBarcelo2003}
C.~Barcelo, S.~Liberati, and M.~Visser.
\newblock Probing semiclassical analogue gravity in bose-einstein condensates
  with widely tunable interactions.
\newblock {\em Phys. Rev. A}, 68:053613, 2003,
  http://arxiv.org/abs/cond-mat/0307491.

\bibitem{SusanaJ.Landau2012}
S.~J. Landau, C.~G. Scoccola, and D.~Sudarsky.
\newblock Cosmological constraints on non-standard inflationary quantum
  collapse models.
\newblock {\em Phys. Rev. D.}, 85:123001, 2012, http://arxiv.org/abs/1112.1830.

\end{thebibliography}

\end{document}